\documentclass[final,times,3p,twocolumn,sort&compress,lefttitle]{elsarticle}

\usepackage{lineno}

\usepackage{amsmath,amssymb,amsfonts}
\usepackage{graphicx}
\usepackage{multirow}
\usepackage{here}
\usepackage{bm}
\usepackage{natbib}
\usepackage{textcomp,nicefrac}
\journal{arXiv}

\usepackage{docmute}

\begin{document}

\begin{frontmatter}

\title{Wide-gap CdTe Strip Detectors for High-Resolution Imaging in Hard X-rays}

\author[ut,kavli]{Shunsaku Nagasawa\corref{cor1}}
\author[ut,kavli]{Takahiro Minami}
\author[isas,kavli]{Shin Watanabe}
\author[kavli,ut]{Tadayuki Takahashi}

\address[ut]{Department of Physics, The University of Tokyo, 7-3-1 Hongo, Bunkyo, Tokyo 113-0033, Japan}
\address[kavli]{Kavli Institute for the Physics and Mathematics of the Universe (Kavli IPMU, WPI), The University of Tokyo, 5-1-5 Kashiwanoha, Kashiwa, Chiba 277-8583, Japan}
\address[isas]{Institute of Space and Astronautical Science, Japan Aerospace Exploration Agency (ISAS/JAXA), 3-1-1 Yoshinodai, Chuo-ku, Sagamihara, Kanagawa 252-5210, Japan}

\cortext[cor1]{Email Address: shunsaku.nagasawa@ipmu.jp}

\begin{abstract}
We propose a new strip configuration for CdTe X-ray detectors, named  ``Wide-gap CdTe strip detector'', in which  the gap between adjacent strips is much wider than the width of each strip.
It has been known that the observed energies of an incoming photon in adjacent strips can be utilized to achieve a position resolution finer than the strip pitch, if and only if the charge cloud induced by an incoming X-ray photon is split into multiple strips and their energies are accurately measured. However, with existing CdTe strip detectors, the ratio of such charge-sharing events is limited.
An idea for a potential breakthrough to greatly enhance the ratio of charge-sharing events is to widen the gaps between strips on the detector.
To test the concept, we developed a wide–gap CdTe strip detector, which has 64 platinum strip electrodes on the cathode side  with some variations in strip pitches from 60 $\mathrm{\mu m}$  (30 $\mathrm{\mu m}$ strip and 30 $\mathrm{\mu m}$ gap width) to 80 $\mathrm{\mu m}$  (30 $\mathrm{\mu m}$ strip and 50 $\mathrm{\mu m}$ gap width). 
We evaluated the performance depending on the strip pitches by irradiating X-rays from ${}^{241}$Am on the detector.
The charge loss due to the wider gaps on the detector was found to be significant to the extent that the assumption that the energy of an incoming photon for a charge-sharing event was the simple sum of the energies detected in adjacent strips lead to a significant degradation in the energy resolution in the accumulated spectrum, compared with those obtained with its predecessor having standard gap-widths.
We then developed a new energy-reconstruction method to compensate for the charge loss.
Application of the method to the data yielded a spectrum with a comparable spectral resolution with that of the predecessor.
The ratio of the charge-sharing events for 17.8 keV events was doubled from that of the predecessor, from 24.3 to 49.9 percent.
We assessed the results with a theoretical simulation based on the Shockley--Ramo theorem and found that they were in agreement. 
\end{abstract}


\end{frontmatter}

\section{Introduction}
Hard X-ray imaging spectroscopy is actively used in the wide field from astronomy to industry and medical imaging. 
However, the imaging capability of existing photon-counting detectors still has a lot of room for improvement.
We have been developing  Double-sided Strip Detectors based on cadmium telluride diode (CdTe-DSD), which provide a far superior energy resolution to conventional detectors, owing to a Schottky junction on the surface of CdTe \cite{takahashi1999high, takahashi2001recent}.

One advantage of the DSD configuration compared with the pixel-type detector is that signals can be measured from both anode and cathode strips. 
Since the $\mu\tau$ products (mobility times lifetime) in CdTe differ by a factor of ten between holes and electrons, the amounts of signals accumulated on the cathode and anode sides differ according to the depth of the interaction point (DoI) \cite{takahashi2001recent, salccin2014fisher}.
If a photon interacts with the detector at a point close to the anode side, for example, more holes are trapped, and the signal on the cathode side, i.e., the total energy, will be weaker than that on the anode side. Conversely, we can estimate the DoI for each event from the difference in the observed signals between the cathode and anode sides and can use it to improve the energy and position resolutions.

Another piece of information provided by the detector in the DSD configuration is charge sharing.
The charge cloud generated in the detector may spread out across multiple strips \cite{koenig2013charge, furukawa2020imaging}, depending on where the incoming photon interacts in the detector.
If charge sharing happens, signals from the event are detected on adjacent strips and the ratio of the amounts of the signals between the strips depends on the location of the original interaction in the detector.
Conversely, utilizing the ratio of the signals enables us to  improve the position resolution to a finer degree than the strip pitch.

Previously, we developed new energy and position reconstruction methods that utilized the DoI estimated from the measured signals from both side strips for 60 $\mu$m strip pitch CdTe-DSDs (strip width of 50 $\mu$m and the gap between adjacent strips of 10 $\mu$m) \cite{furukawa2020imaging}. The detector is developed for the Focusing Optics X-ray Solar Imager (FOXSI-3) sounding rocket experiment \cite{krucker2009focusing, musset2019ghost}.
We achieved an energy resolution of 0.8 keV (FWHM) for the 13.9 keV line peak using both sides of energy information for single and double-strip events (the type of events of which the signal is split to and detected on two adjacent strips).
We also demonstrated by modeling the charge propagation on the basis of the DoI and sharing energies between adjacent strips that the position resolution finer than the strip pitch could be achieved.
However, the ratio of charge-sharing events for 20 keV photons on the cathode side was limited to 25\% \cite{furukawa2020imaging}.
Given that charge sharing is essential to achieve a sub-strip position resolution, a higher ratio is more desirable to achieve a better position resolution.
In particular, the position resolution of $\lesssim 30~\mathrm{\mu m}$ and the energy resolution of $\lesssim 1~\mathrm{keV}$ are required at the same time for the upcoming fourth launch FOXSI-4 \cite{2020AGUFMSH0480011G, Camilo2021} in 2024 to provide simultaneous diagnostics of specifically and spatially separated coronal and chromospheric emission.

We propose a novel concept of widening a gap between adjacent strips to enhance the ratio of charge-sharing events.
One way to increase the charge-sharing probability is to reduce a strip pitch, which we have attempted with some success in the past \cite{furukawa2020imaging, furukawa2019development}. 
However, it is difficult to pursue this direction further to ensure a certain yield level with the current technology.
This concept also has merit in making a sensitive area of the detector larger while keeping the number of read-out channels.
So far, the concept has never been experimentally proved, and it may have downsides, such as potential degradation in spectral performance and/or detection efficiency.

To examine the concept, we developed a wide-gap CdTe strip detector, in which the gap between adjacent strips is much wider than the width of the strips. In this paper, we report the results of our experiments to  evaluate  the effect of strip and gap widths on the charge-sharing properties, as well as 
spectral performance and detection efficiency. We also describe our new energy-reconstruction method to compensate for the charge loss caused by wide gaps.
Finally, we  report the results of a theoretical simulation to evaluate our experimental results. 

\begin{figure}[H]
\begin{center}
\includegraphics[width=1.0\hsize]{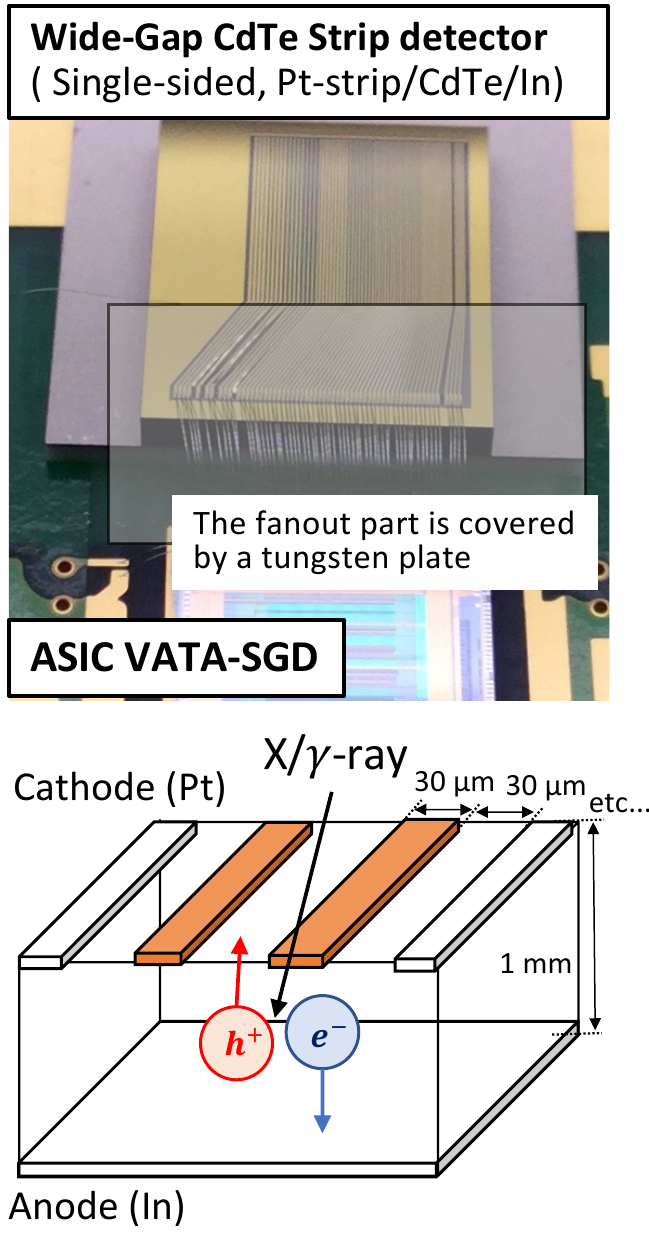}
\caption{A photograph of the wide-gap CdTe strip detector (upper) and a schematic view of the electrode configuration (lower). The detector is connected to a VATA-SGDs ASIC \cite{watanabe2014si} for pulse-height measurement. The fanout part is covered with a 1 mm thick tungsten plate to accurately evaluate the difference in charge-sharing properties caused by strip and gap widths.} 
\label{setup_mini}
\end{center}
\end{figure}

\section{Detector configurations}
Figure \ref{setup_mini} shows a photograph of the wide-gap CdTe strip detector that we developed and a schematic view of the electrode configuration.
The detector is 1 mm thick and has 64 platinum strip electrodes on the cathode side and a common indium electrode on the anode side (i.e., The detector is a single-sided strip detector configuration).
The widths of strips and gaps between the strips are varied from $60~\mathrm{\mu m}$ strip pitch ($30~\mathrm{\mu m}$ strip width and $30~\mathrm{\mu m}$ gap width) to $80~\mathrm{\mu m}$ strip pitch ($30~\mathrm{\mu m}$ strip width and $50~\mathrm{\mu m}$ gap width).
The signal is read only from the cathode side with the Application Specific Integrated Circuits (ASICs) (i.e., the energy and position (1-dimension) information is obtained only from the cathode side.).
The ASIC has 64 channels, and each channel contains a preamplifier, a fast shaper for self-triggering, and a slow shaper for pulse-height measurement (VATA-SGD ASICs \cite{watanabe2014si}, originally developed for the Soft Gamma-ray Detector (SGD) onboard the ASTRO-H (Hitomi) satellite \cite{takahashi2014astro}).

The detector was housed in a metal box and kept cooled at $-10$ degrees Celsius.
The bias voltage was set to 500 V.
To accurately evaluate the difference due to strip and gap widths, we covered the fanout part (which connects the detector and ASIC) with a 1 mm thick tungsten plate and irradiated the detector  on the cathode side with X-rays from  an ${}^{241}$Am source. The distance between the detector and the radiation source was set to $D = 120~\mathrm{mm}$ for the size of the detector $L = 4.95~\mathrm{mm}$ and accordingly,
the difference in X-ray intensity due to the difference in the distance from the radiation source  to the detector surface was limited to ${(L/D)}^2 = {(4.95~\mathrm{mm}/120~\mathrm{mm})}^2 \sim 0.2\% $

In the following sections, we present the results based on signals extracted from five consecutive channels in one or more sets out of the following four sets of combinations for strip and gap widths.
We focus on the energy range for the FOXSI observation target ($E \lesssim 30~\mathrm{keV}$).
\begin{description}
\setlength{\parskip}{0cm}
\setlength{\itemsep}{0cm}
\item[Set (A)]~~strip width 30~$\mathrm{\mu m}$ -- gap width 30~$\mathrm{\mu m}$
\item[Set (B)]~~strip width 30~$\mathrm{\mu m}$ -- gap width 40~$\mathrm{\mu m}$
\item[Set (C)]~~strip width 30~$\mathrm{\mu m}$ -- gap width 50~$\mathrm{\mu m}$
\item[Set (D)]~~strip width 40~$\mathrm{\mu m}$ -- gap width 40~$\mathrm{\mu m}$
\end{description}

\begin{figure*}[!]
\begin{center}
\includegraphics[width=1.0\hsize]{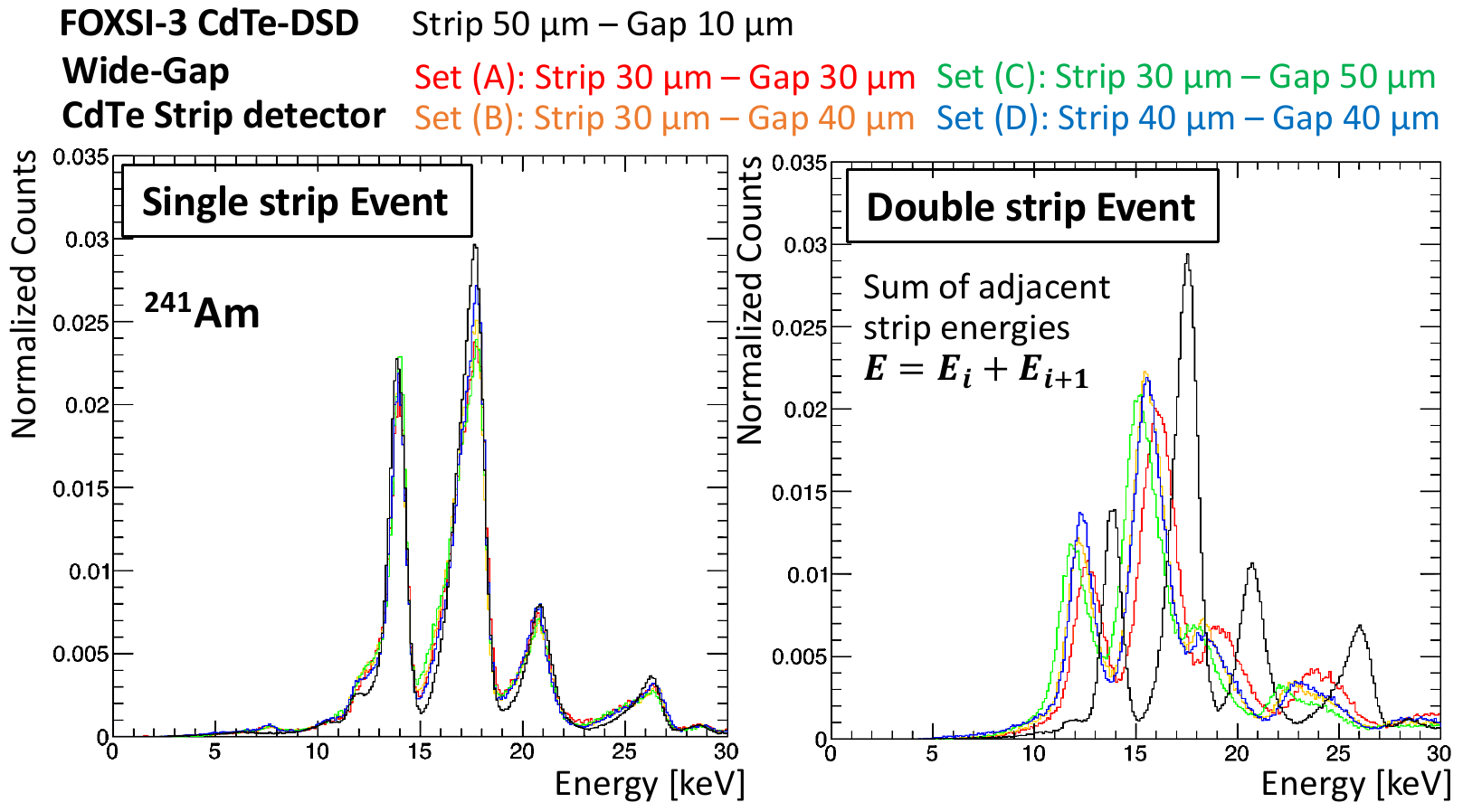}
\caption{Spectra of ${}^{241}$Am for (left) single-strip events and (right) double-strip events in each set of strip and gap regions: sets (A)--(D) in red, orange, green, and blue, respectively. The energy spectrum of FOXSI-3 CdTe-DSD is also overlaid in black. For a double-strip event, the sum of the strip energies is adopted as its photon energy (i.e., $E=E_i+E_{i+1}$, where $E_i$ and $E_{i+1}$ are the detected energies on the $i$-th and ($i+1$)-th strips, respectively). The energy threshold of each strip is set to 1.5 keV.}
\label{spect_before}
\end{center}
\end{figure*}

\begin{figure*}[!]
\begin{center}
\includegraphics[width=1.0\hsize]{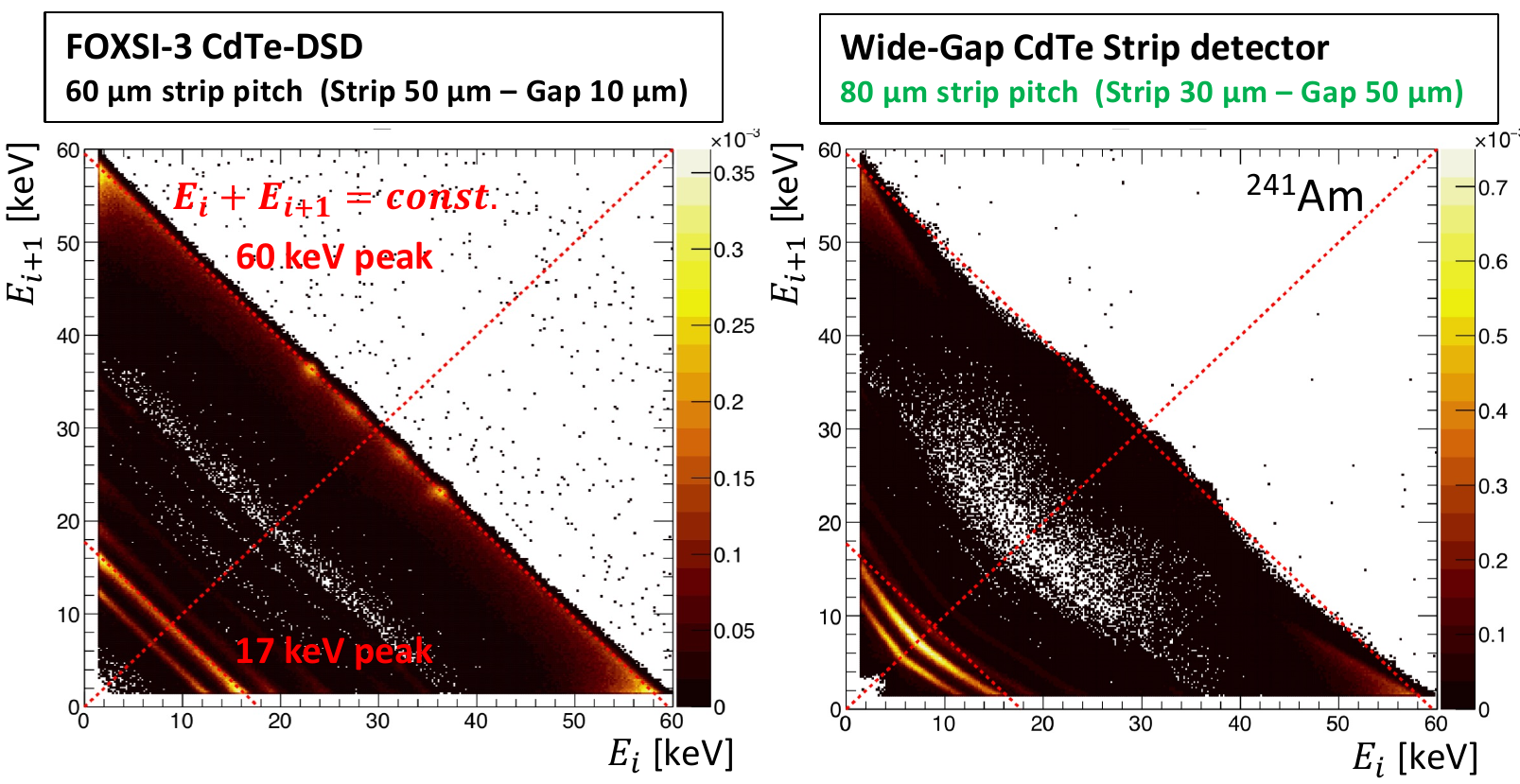}
\caption{Relationships between the energies detected on adjacent strips for (left) FOXSI-3 CdTe-DSD and the (right) wide-gap detector of the set (c) for double-strip events. The four points in the 60 keV peak line are the X-ray fluorescence of Cd and Te (23 keV and 27 keV). The 60 keV events are less for the wide-gap detector of the set (c) because the ratio of three and four strip events are increased to $\sim 50$\%.}
\label{ratio_energy}
\end{center}
\end{figure*}

\begin{figure*}[!]
\begin{center}
\includegraphics[width=1.0\hsize]{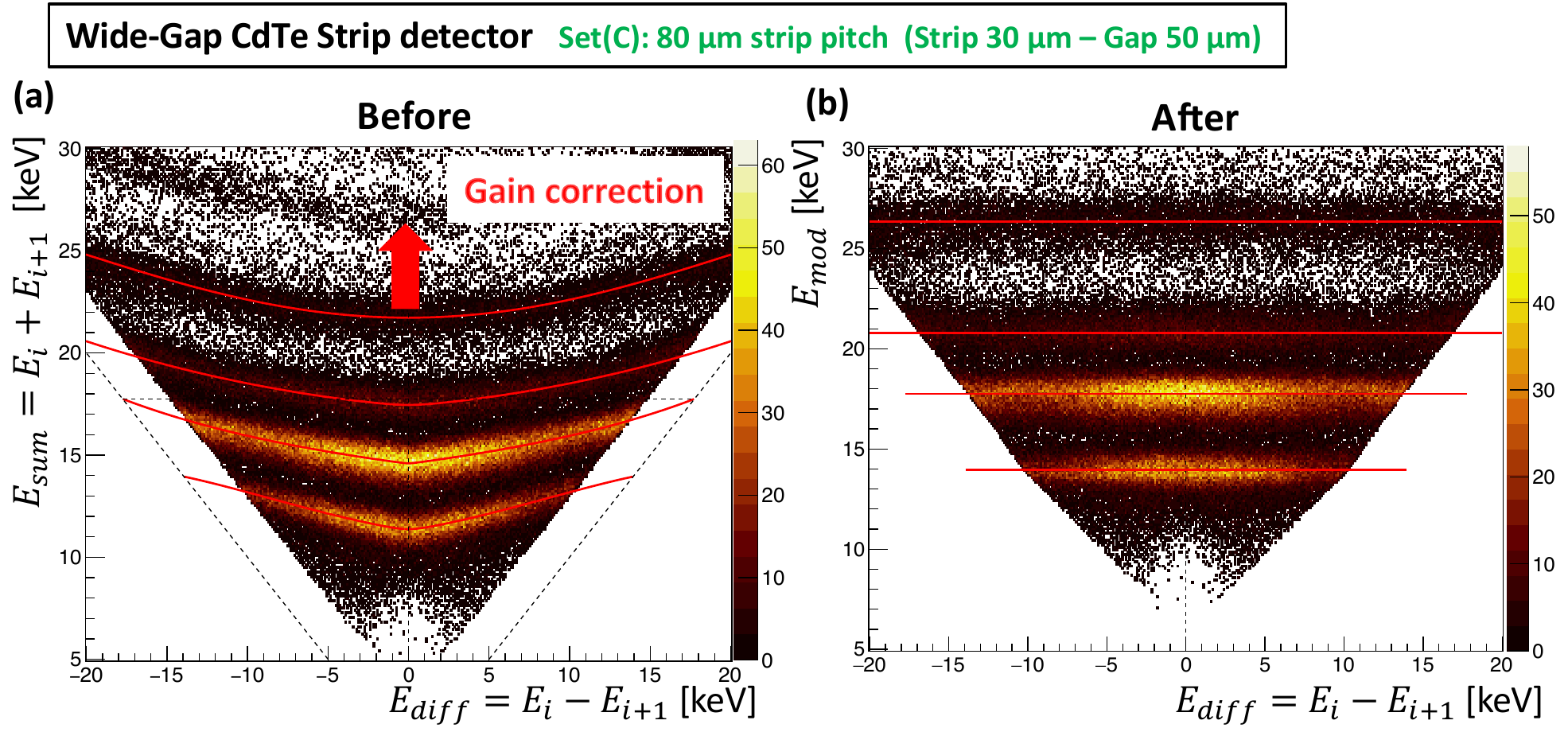}
\caption{(a) Relationships between the difference $ E_{diff} = E_{i} - E_{i+1}$ and the sum $E_{sum} = E_i + E_{i+1}$ of the strip energies for double-strip events. The peak positions of the V-shaped structures are fitted  with quartic functions.
(b) the gain correction factors derived from the determined quartic model functions are multiplied to panel (a), resulting in the relationship where  the reconstructed energy $E_{mod}$ is a constant value with respect to $E_{diff}$.}
\label{mod_method}
\end{center}
\end{figure*}

\begin{figure*}[!]
\begin{center}
\includegraphics[width=1.0\hsize]{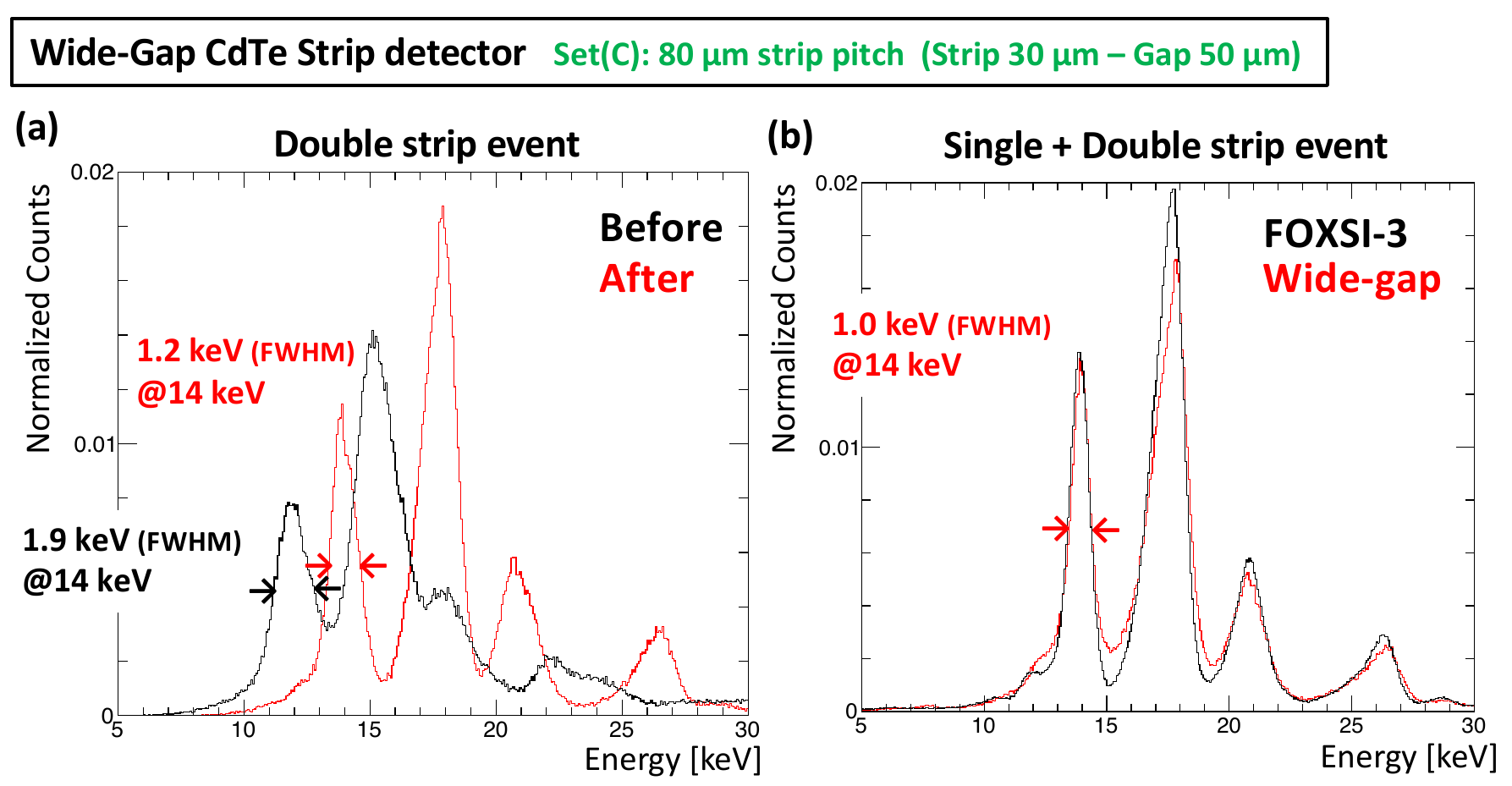}
\caption{(a) Spectra of the (black) sum of the strip energies $E_{sum}$ and the (red) reconstructed energy $E_{mod}$ for double-strip events. All $E_{diff}$ events are used. (b) A spectrum for single-strip and reconstructed double-strip events (red). The spectrum of FOXSI-3 CdTe-DSD is also overlaid (black).}
\label{reconst_spect}
\end{center}
\end{figure*}

\section{Effect on Spectroscopic Performance}\label{sec:spec}
\subsection{Energy loss in wide-gap configurations}
Figure \ref{spect_before} shows the energy spectra of ${}^{241}$Am source obtained  with each set of strip-and-gap-widths regions. Here in the spectrum, the sum of the strip energies $E_{sum}$ is used for a double-strip event.
Let us define $E_i$ and $E_{i+1}$ as the strip energies directly converted from the signals detected in the adjacent $i$-th and ($i+1$)-th strips and then $E_{sum}=E_i+E_{i+1}$. The energy threshold of each strip is set to 1.5 keV. The events spreading over three or more strips accounted for the remaining 2\%, and we would ignore these events in the following analyses.
In Figure~\ref{spect_before}, we also overlay, for comparison, the energy spectrum of FOXSI-3 CdTe-DSD with a 60 $\mu$m strip pitch  (50 $\mu$m strip and 10 $\mu$m\ gap), the thickness of which is 0.75 mm, where the bias voltage is set to 200 V.

The obtained spectra for single-strip events show no significant deterioration in spectral performance in the wide-gap CdTe detectors (0.9 keV in FWHM for the widest-gap set (C) at the 13.9 keV peak) from that in FOXSI-3 CdTe-DSD, except for the presence of a low-energy tail component, which appears increasingly more noticeable for wider-gap detectors.

In contrast, the spectral performance for double-strip events apparently deteriorated with the peaks shifting to the lower-energy side. 
Figure~\ref{ratio_energy} shows the relationships of the energies detected in adjacent strips for double-strip events. 
For FOXSI-3 CdTe-DSD, there is a linear relationship between the energies detected in adjacent strips ($E_i+E_{i+1}=const.$).
This implies that a simple summation of the energies on adjacent strips is a reasonable method to reconstruct incident X-ray energies.
In contrast, for wider-gap detectors, the sum of the energies deviates from the constant and  is smaller. The tendency is particularly noticeable for the events for which the energies between adjacent strips are similar ($E_i \simeq E_{i+1}$); the events of this type occur when  photons interact with the material in the detector at a location  close to the center of the gap. The spectrum in Figure~\ref{spect_before} and relationship plot in Figure~\ref{ratio_energy} indeed imply that part of the charge induced by an incoming photon is lost during charge accumulation in the detector and that the amount of the charge loss is likely to be a function of the photon-interaction position inside the detector \cite{koch2021charge,abbene2018digital}. 

In the next subsection, we develop an energy reconstruction method for compensating for the charge loss as a result of an adoption of wide gaps on the detector, with which the energy dependence of charge-sharing properties is accurately assessed, with the aim of achieving the similar spectral performance to that with predecessors (specifically, FOXSI-3 CdTe-DSD).

\subsection{Energy Reconstruction algorithm and improved spectra}
The amount of energy loss due to the wide-gaps depends on the difference in the energies in adjacent strips, which depends on the interaction positions of photons. 
We developed an energy-reconstruction method by empirically determining the relationships of the sum ($E_{sum} = E_i + E_{i+1}$) and difference ($E_{diff} = E_{i}-E_{i+1}$) of the strip energies for double-strip events. 
In this section, we focus on the widest-gap set (C).

First, we fit the peak position of the V-shaped structure of each ${}^{241}$Am line peak with a quartic function ($a_0 + a_1|E_{diff}| + a_2E_{diff}^2 + a_4E_{diff}^4$, see Figure \ref{mod_method}(a)). 
Second, we obtain the reconstructed energy $E_{mod}$ by multiplying the sum of the strip energies of a double-strip event $E_{sum}$ by the gain correction factor (see Figure \ref{mod_method} (b)).
The gain correction factor is calculated based on the best-fit model functions for each line peak of ${}^{241}$Am so that $E_{mod}$ is a constant value with respect to $E_{diff}$.
Finally, we calculate the reconstructed energies between the line peaks of ${}^{241}$Am with the linear interpolation using the Delaunay triangulation \cite{delaunay1934bn, lee1980two}, which utilizes the TGraph2D class in the ROOT framework \cite{brun1997root}.

Figure \ref{reconst_spect} (a) shows the spectra of the sum of the strip energies $E_{sum}$ and the reconstructed energy $E_{mod}$ for double-strip events. 
The deterioration of spectroscopic performance due to the charge loss is properly reconstructed, and the energy resolution is improved from 1.9 keV to 1.2 keV (FWHM) at the 13.9 keV peak after the reconstruction. 
Figure \ref{reconst_spect} (b) shows the summed spectrum of the single-strip events and the reconstructed double-strip events (51\% and 47\% of the total events, respectively).
The energy resolution for the combined spectrum of single-strip and reconstructed double-strip events was found to be 1.0 keV (FWHM) at the peak of 13.9 keV, while a sufficient spectroscopic performance is maintained despite wider gaps than those of the FOXSI-3 CdTe-DSDs.

Let us give a brief comment on the limitation of the current detector for the pilot study and potential future improvement.
In our current wide-gap detector, signals were read only from the cathode side. Ideally, a wide-gap structure can be implemented also on the anode side, and signals are also read from it; then, we can, in principle, take into account the effect of charge trapping as a function of the DoI by using energy information obtained from both the anode and cathode sides \cite{furukawa2020imaging}, although the reconstruction method requires some further development to accommodate the information of DoI.
As such, this detector and method have a realistic potential for further improvement.

\begin{figure}[H]
\begin{center}
\includegraphics[width=1.0\hsize]{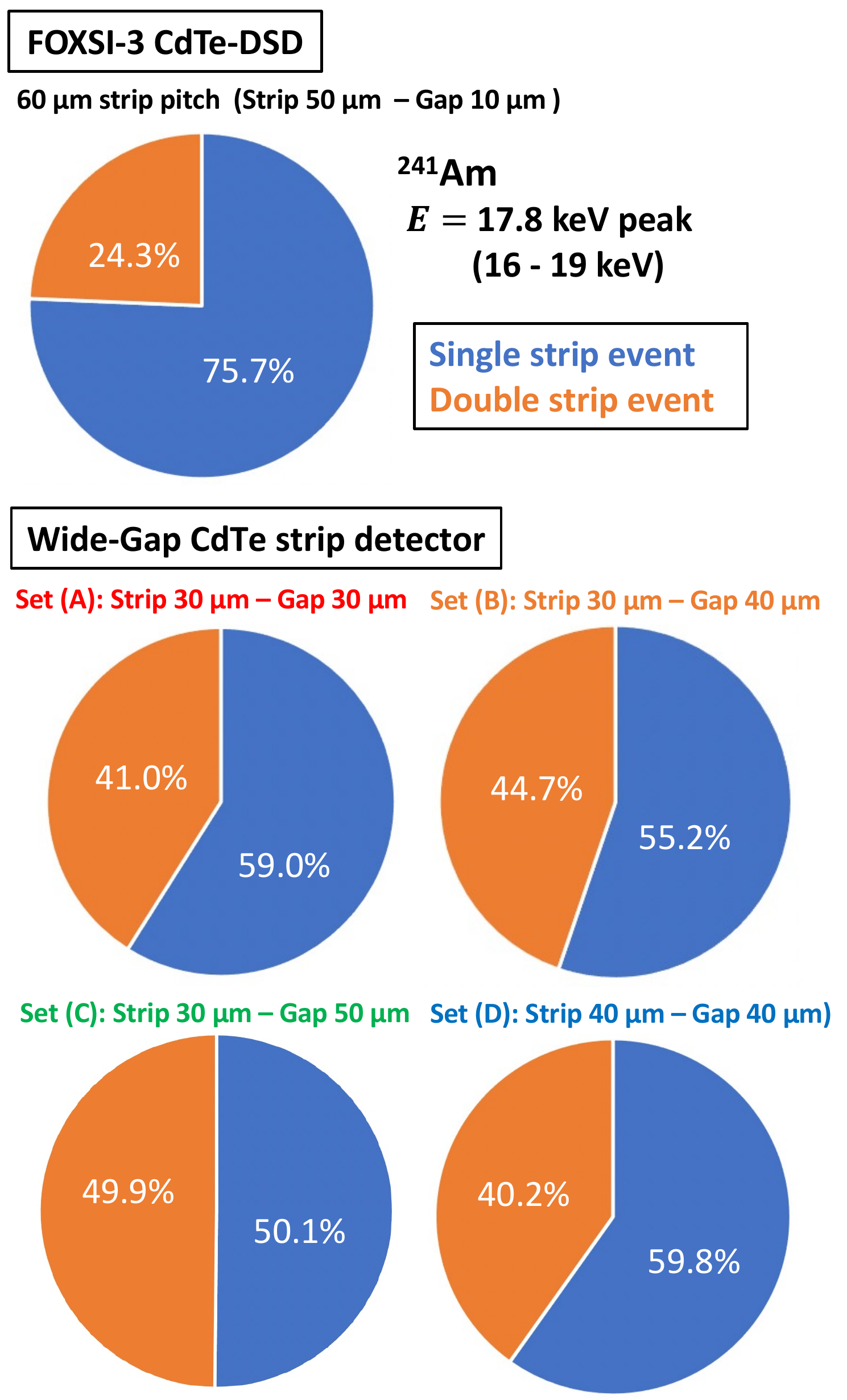}
\caption{Ratio of charge-sharing events for each set of strip-and-gap widths where the events corresponding to the 17.8 keV peak of ${}^{241}$Am are extracted. For comparison, the ratio of charge-sharing events for  FOXSI-3 CdTe-DSD is also shown. The energy threshold of each strip is set to 1.5 keV.
}
\label{percent}
\end{center}
\end{figure}

\section{Ratio of Charge Sharing}
By developing the energy reconstruction method to compensate the charge loss by wide–gaps, it makes possible to accurately assess the energy dependence of charge sharing properties.
Figure \ref{percent} shows the obtained ratios of charge-sharing events among the events for the incoming photons at the 17.8 keV peak of ${}^{241}$Am, or specifically, in an energy range of 16--19 keV, for FOXSI-3 CdTe-DSD and our sets (A)--(D) of wide-gap CdTe-DSD.
The energy threshold of each strip was set to 1.5 keV.

In the case of the same 60 $\mathrm{\mu}$m strip pitch, the ratio of double-strip events for FOXSI-3 CdTe-DSD (gap 10~$\mathrm{\mu}$m) was 24.3\%, whereas the ratio increases to 41.0\% when the gap width is widened to 30~$\mathrm{\mu}$m (set (A)). 
Moreover, as the gap width is widened to 40~$\mathrm{\mu}$m and 50~$\mathrm{\mu}$m, the ratio of double-strip events increases  up to 49.9\% for the widest-gap set (C).  In consequence, we confirmed that the ratio of double-strip events can be increased by increasing the gap width.

\section{Detection Efficiency}
We evaluated the detection efficiency of the wide-gap detector, which might be degraded from the existing model. Table \ref{det_eff} shows the count rates of events for an energy range of 10--30 keV  from ${}^{241}$Am. 
The error of the count rate is evaluated by dividing the total 154 hours of data by 1 hour and taking the standard deviation of each count rate.
In the analysis, events from five consecutive strips were extracted for each set of strip and gap widths. In this case, if the detection efficiency on the detector surface and the spatial distribution of the irradiation from the ${}^{241}$Am source are both uniform, the count rate is proportional to the strip pitch.  Table~\ref{det_eff} also lists the count rates normalized by the strip pitch for comparison.
We found that the detection efficiency was invariable in the 10--30 keV energy band regardless of the gap width, even as large as 50~$\mathrm{\mu}$m within the margin of error.
\begin{table}[htb]
  \centering
  \caption{Count rates at each set of strip and gap widths in the energy range of 10--30 keV of X-rays from ${}^{241}$Am}
  \scalebox{0.8}{
  \begin{tabular}{l|c|c}
     & Count Rate &  Normalized Count Rate \\
    & [counts/s] & [$10^{-3}$ counts/s/$\mathrm{\mu m}$] \\\hline \hline
    Set (A):& \multirow{2}{*}{$0.345 \pm 0.020$} & \multirow{2}{*}{$5.75 \pm 0.34$} \\
    strip 30~$\mathrm{\mu}$m -- gap 30~$\mathrm{\mu}$m & & \\
    \hline
    Set (B):& \multirow{2}{*}{$0.424 \pm 0.022$} & \multirow{2}{*}{$6.05 \pm 0.31$} \\
    strip 30~$\mathrm{\mu}$m -- gap 40~$\mathrm{\mu}$m  & & \\
    \hline
    Set (C):& \multirow{2}{*}{$0.488 \pm 0.027$} & \multirow{2}{*}{$6.10 \pm 0.33$} \\
    strip 30~$\mathrm{\mu}$m -- gap 50~$\mathrm{\mu}$m & & \\
    \hline
    Set (D):& \multirow{2}{*}{$0.492 \pm 0.026$} & \multirow{2}{*}{$6.14 \pm 0.33$} \\
    strip 40~$\mathrm{\mu}$m -- gap 40~$\mathrm{\mu}$m & & \\
    \hline
  \end{tabular}
  }
  \label{det_eff}
\end{table}

\section{Modeling of Detector Response}
We conducted the response simulation to make a prospective of the position resolution we can obtain for the wide-gap CdTe strip detector. 
Accurate modeling of the X-ray detector response is also important for accurate interpretation of the results of experiments with the detector and future detector development.
We numerically investigated the response of the wide-gap CdTe strip detector, specifically the dependence of charge-sharing properties on photon-interaction positions, based on the Shockley--Ramo theorem that an induced charge at electrodes is calculated by integrating a weighting potential along a carrier trajectory.

\subsection{Simulation setup}
The charge $Q$ induced on electrodes of the detector can be calculated  according to the Shockley--Ramo theorem \cite{eskin1999signals, he2001review}. 
In the theorem, a mathematical concept of ``weighting potential'' ~$\phi_{w}$ is introduced. 
When the carrier density $q(\vec{r})$  moves from  the interaction position $\vec{r_0}$ to the position $\vec{r_f}$, the induced charge $Q$ at the electrode on the cathode or anode side is  given by: 

\begin{align}
Q=\int^{\vec{r_f}}_{\vec{r_0}} q(\vec{r'})\vec{E_w}(\vec{r'})\cdot d\vec{r'},
\label{chargeeff}
\end{align} 
where $\vec{E_w} = -\nabla \phi_w$ is the weighting field.
The trajectory of the carrier is a function of the actual electric field $\vec{E}$.

\begin{figure}[htb]
\begin{center}
\includegraphics[width=1.0\hsize]{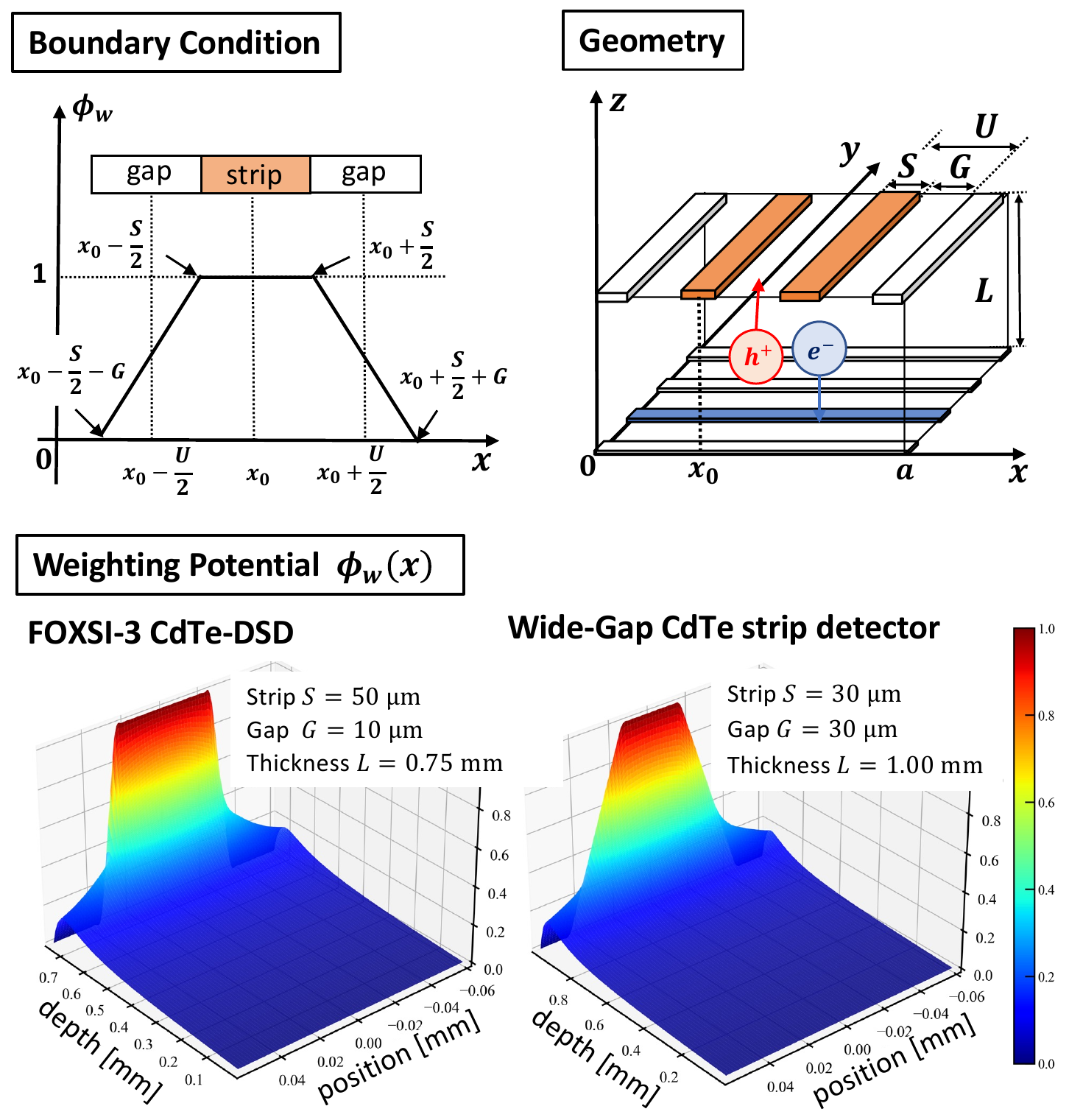}
\caption{Upper:  Schematic illustration of the boundary condition and  detector geometry for the simulation. Lower: Weighting potential for FOXSI-3 CdTe-DSD and the wide-gap CdTe strip detector (set (A)). }
\label{weight_pot}
\end{center}
\end{figure}

The weighting potential $\phi_{w}$  depends only on the detector geometry and electrode configuration and can be calculated by solving Laplace's equation $\Delta \phi_{w} = 0$ with  boundary conditions of $\phi_{w} = 1$ at the electrode of interest and $\phi_{w} = 0$ at all other electrodes. As for the boundary condition at the gaps, we assumed that the potential on the gap between two electrodes  is a linear decreasing function from 1 at the edge of the electrode of interest to 0 at the edge of the  other electrode.
Figure~\ref{weight_pot} (upper panel) illustrates the boundary conditions and the detector geometry. The weighting potential $\phi_{w}$ is calculated as follows:
\begin{flalign}
    &\mbox{Anode: }\nonumber	\\
    &~~\phi_{w} = \sum _{m=1}^\infty A_m(x_c) \sin{\left(\frac{m\pi}{a}x\right)}\sinh{\left(\frac{m\pi}{a}z\right)}\\
    &\mbox{Cathode: }\nonumber \\
    &~~\phi_{w} = \sum _{m=1}^\infty A_m(y_c) \sin{\left(\frac{m\pi}{a}y\right)}\sinh{\left(\frac{m\pi}{a}(L-z)\right)}\\
    &~A_m(x_{c}) = \frac{8a}{(m\pi)^2 G \sinh{\left(\frac{m\pi L}{a}\right)}} \nonumber\\ 
    &~~~~~~~~~\times \sin{\left(\frac{m\pi}{a}x_{c}\right)}\sin{\left(\frac{m\pi}{a}\frac{U}{2}\right)}\sin{\left(\frac{m\pi}{a}\frac{G}{2}\right)},
\end{flalign}
\label{weighting_pot2}
where $a$ and $L$ are the size and thickness of the detector, $U$ and $G$ are the widths of the strip pitch and gaps, respectively, and a pair of $x_c$ and $y_c$ is the center position of the target electrode of calculation \cite{hagino2012imaging}. 
Figure \ref{weight_pot} (lower panel) shows the calculated weighting potential $\phi_{w}$ for FOXSI-3 CdTe-DSD and the wide-gap CdTe strip detector (set (A)). 
The potential distribution was found to show a steep rise  near the electrode; the fact suggests that  the signal  mainly  originates from  carrier movement near the read-out electrode and that the motions of holes and electrons mainly contribute to the signals on the cathode and anode sides, respectively (``small-pixel effect'').
In addition, the weighting potential in the gap region has a higher value for the wide-gap detector.
Hence, a considerable amount of signal should be induced in also the adjacent strip, in addition to the strip closest to the interaction point.

For the sake of simplicity, we assumed that the electric field was uniform ($E = V/L$, parallel to $z$-direction).
The number of strips is set to 128 on both sides, and the strip and gap widths are set in common (50~$\mathrm{\mu m}$ strip and 10~$\mathrm{\mu m}$ gap width for FOXSI-3 CdTe-DSD, and 30~$\mathrm{\mu m}$ strip and 50~$\mathrm{\mu m}$ gap width for the wide-gap detector (set(c))), and the effects of a guard ring and a fanout part are not included.
We also assumed that the density of the carriers generated by incoming X-rays at the interaction depth $z_i$ was expressed with regard to the number of electron-hole pair productions $n$:
\begin{align}
q_{e,h}(t) = (\mp e)n\exp\left(-\frac{|z(t)-z_i|}{\lambda_{e,h}}\right), 
\end{align}
where $\lambda_{e,h} = (\mu\tau)_{e,h}E$ is the mean free path of carriers in CdTe. The carriers are decreased because of charge trapping.
In the calculation, we used typical values of $\left(\mathrm{\mu\tau}\right)_e, \left(\mathrm{\mu\tau}\right)_h =2.0\times10^{-3}, 1.2\times10^{-4}~$cm$ ^2$/s, respectively.
We also convoluted the response of pulse shaping in ASIC according to the following form:
\begin{align}
V_{out}(t)\propto \int^{t}_{0} h(t-t')\frac{dQ}{dt'}dt'.
\label{shaper}
\end{align}
We assumed that the induced current $dQ/dt$ was regarded as the impulse current and that $h(t)$ was a semi-Gaussian given by $h(t) = (t/\tau)e^{t/\tau}$ ($\tau = 3~\mathrm{\mu s}$).
The maximum value of output voltage $V_{max}$ was assigned to the output pulse height.
The output pulse height $V_{max}$ depends on the interaction depth $z_i$.
Given that the dependence of interaction depth is smaller in the FOXSI target energy range ($E \lesssim 30~\mathrm{keV}$ with the penetration length $l$ of CdTe for 30 keV X-rays being $l\sim 0.1~\mathrm{mm}$),
we took the weighted average of the output pulse height over the interaction depth and then converted it to detected energy in each strip:
\begin{align}
\overline{V_{max}}= \frac{\int^{L}_{0}V_{max}(z_i)e^{-\frac{z}{l}}dz_i}{\int^{L}_{0}e^{-\frac{z_i}{l}}dz_i}.
\label{pulseheight}
\end{align}

\begin{figure*}[htb]
\begin{center}
\includegraphics[width=0.95\hsize]{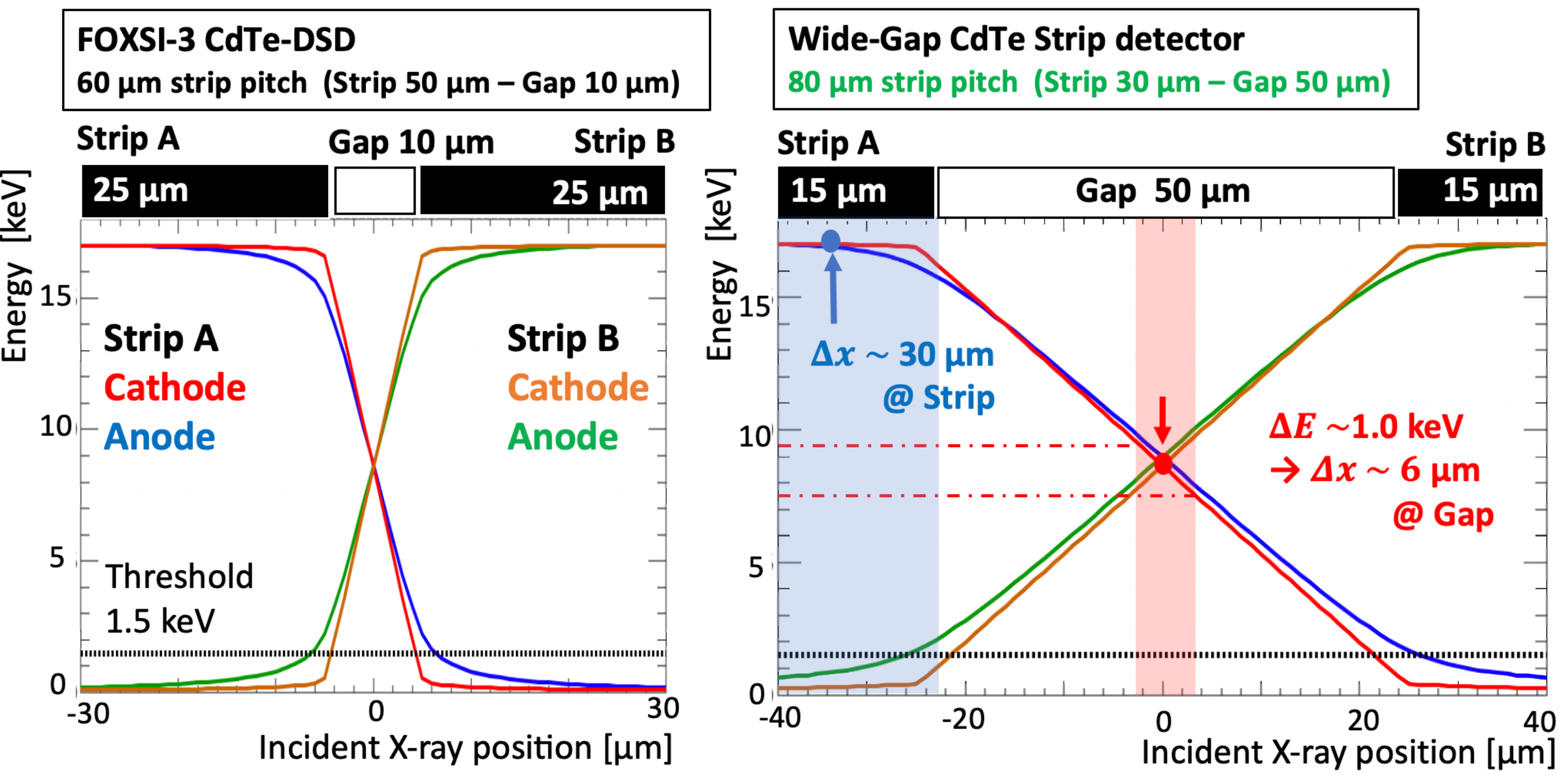}
\caption{Relationships between the detected energy in two adjacent strips A and B and the incident X-ray position from the center of strip A to the center of the other strip B.
Red and blue represent the energies obtained from the cathode and anode sides, respectively,  detected on strip A, and 
 orange and green represent  those detected in strip B.
The X-ray energy is set to 17.8 keV and the threshold is set to 1.5 keV.
}
\label{check_posres}
\end{center}
\end{figure*}

\subsection{Results}
First, we calculated the ratio of double-strip events on the cathode side  from uniformly irradiated monochromatic X-rays  at 17.8 keV. The energy threshold was set to 1.5 keV. 
Table \ref{modeling_result} shows the results of the simulation. 
We found that the simulation results were consistent with our experimental results within $\sim$10\% for FOXSI-3 CdTe-DSD and $\sim$4\% for the wide-gap detector.
We conjecture that the inconsistency would be mitigated if the effects of diffusion of the carrier density and non–uniformity of the electric field are taken into account.
The detail is beyond the scope of this work, but the effect of charge diffusion may be important for the FOXSI-3 CdTe-DSD at least.
It is demonstrated for the FOXSI-3 CdTe-DSD that charge-sharing events at the cathode side increase as photons interact near the surface of the anode side \cite{furukawa2020imaging}.
In addition, based on the synchrotron beam scanning test \cite{duncan2022modeling}, the typical width of a ``zone" where double-strip events are the most common in the gap region is estimated to $\sim 20~\mathrm{\mu m}$, which is slightly large compared with the gap width of 10 $\mathrm{\mu m}$.
These results can be caused by charge diffusion. 

\begin{table}[htb]
  \centering
  \caption{Comparison between the simulation and experimental results}
  \scalebox{0.76}{
  \begin{tabular}{l|c|c|c|c}
      & \multicolumn{2}{c}{Simulation} &\multicolumn{2}{|c}{ Experiment}\\\cline{2-5}
      Cathode side ($E=17.8~\mathrm{keV}$)&Single&Double&Single&Double\\\hline\hline 
    FOXSI-3 CdTe-DSD 
    & \multirow{2}{*}{85.5 \%} & \multirow{2}{*}{14.5 \%}  & \multirow{2}{*}{75.7 \%}  & \multirow{2}{*}{24.3 \%}\\
    strip 50 $\mathrm{\mu}$m -- gap 10 $\mathrm{\mu}$m & & & &  \\\hline
    Wide-gap strip detector 
    & \multirow{2}{*}{46.3 \%} & \multirow{2}{*}{53.7 \%}  & \multirow{2}{*}{50.1 \%}  & \multirow{2}{*}{49.9 \%}\\
    strip 30 $\mathrm{\mu}$m -- gap 50 $\mathrm{\mu}$m & & & &  \\\hline
  \end{tabular}  
  }
  \label{modeling_result}
\end{table}

Next, we calculated the ratio of the strip energies between two adjacent strips for a given incident X-ray position for FOXSI-3 CdTe-DSD and the wide-gap strip detector (set(c)), where the X-ray energy was set to 17.8 keV and the threshold was set to 1.5 keV.
Figure \ref{check_posres} shows the detected energy in two adjacent strips A and B for the given incident X-ray position from the center of strip A to the center of the next strip B. 
For FOXSI-3 CdTe-DSD, the detected energy  was dominated on one side of the strip and the detected energy steeply shifted to the next strip in the gap region.
By contrast, for the wide-gap detector, the detected energy  was found to change gradually from one side of the strip to the other, depending on the incident X-ray position. 
The obtained relationship implies that if we could determine the anode and cathode side of energies with an accuracy of 1 keV, 
a position resolution  at the gap position is expected to be $\sim 6~\mathrm{\mu m}$,  
as derived through evaluation of the energies on adjacent strips. In addition, even for single-strip events at the strip position, a position resolution of $\sim 30~\mathrm{\mu m}$ can be achieved.

\section{Conclusion}
We propose a novel concept of widening a gap between adjacent strips to enhance the ratio of charge-sharing events and achieve a position resolution finer than the strip.
This concept also has merit in making a sensitive area of the detector larger with the limited read-out channels.
To examine the concept, we developed a wide–gap CdTe strip detector and evaluated the effect of strip and gap widths on the charge-sharing proprieties.
We obtained an energy resolution of 1.0 keV in FWHM at 13.9 keV by applying a correction to compensate for the charge loss caused by wide gaps. 
The correction also makes it possible to assess the energy dependence of the charge-sharing property accurately.
We found that the ratio of double-strip events could be increased by $\sim 50\%$ on the cathode side by making the gap wider. 
Our simulation work suggests that the position resolution is expected to be $\sim 6~\mathrm{\mu m}$  at the gap position and to be $\sim 30~\mathrm{\mu m}$ even for single-strip events at the strip position. 
Although, in this study, the evaluation test is conducted using the single-sided strip detector (signals were read only from the cathode side), we can make a prospective that a sub–strip position resolution is well achievable with wide-gap CdTe-DSDs, while obtaining a sufficient spectroscopic performance of $\sim$1 keV for the upcoming FOXSI-4 flight.
Furthermore, the performance of wide-gap CdTe-DSDs has a realistic potential for further improvement by taking into account the DoI effect using both anode and cathode sides of energy information.

\section*{Acknowledgement}
This work was supported by JSPS, Japan KAKENHI Grant Numbers 18H05457, 20H00153, and 22J12583.
S.N. is also supported by FoPM (WINGS Program) and JSR Fellowship, the University of Tokyo, Japan.

\bibliography{main}

\end{document}